\documentclass[preprints,article,accept,moreauthors,pdftex,10pt,a4paper]{mdpi} 
\usepackage{natbib}            
\firstpage{1} 
\articlenumber{x}
\doinum{10.3390/------}
\pubvolume{xx}
\pubyear{2018}
\copyrightyear{2018}
\history{Received: date; Accepted: date; Published: date}

\newcommand{\gtrsim}{\stackrel{\textstyle >}{_\sim}}
\newcommand{\lesssim}{\stackrel{\textstyle <}{_\sim}}

\Title{NEUTRINO EMISSIVITY IN THE QUARK-HADRON MIXED PHASE}

\Author{William M. Spinella$^{1}$, Fridolin Weber$^{2,3*}$,
  Milva G. Orsaria$^{4}$ and Gustavo A. Contrera$^{5}$}

\AuthorNames{William M. Spinella, Fridolin
  Weber, Gustavo A. Contrera, and Milva G. Orsaria}

\address{$^{1}$ \quad Department of Sciences, Wentworth Institute of
  Technology, 550 Huntington Avenue Boston, MA 02115, USA;
  spinellaw@wit.edu \\
$^{2}$ \quad Department of Physics, San Diego State University, San Diego, CA
  92182, USA; fweber@sdsu.edu \\
$^{3}$ \quad University of California at San Diego, La Jolla, CA
  92093, USA; fweber@ucsd.edu \\
$^{4}$ \quad Grupo de Gravitaci\'on, Astrof\'isica y Cosmolog\'ia,
  Facultad de Ciencias Astron{\'o}micas y Geof{\'i}sicas, Universidad
  Nacional de La Plata,
  La Plata,
  Argentina; CONICET,
  Buenos Aires,
  Argentina; morsaria@fcaglp.unlp.edu.ar\\ 
$^{5}$ \quad IFLP, UNLP, CONICET, Facultad de Ciencias Exactas,
  La Plata, Argentina; Grupo de Gravitaci\'on, Astrof\'isica
  y Cosmolog\'ia, Facultad de Ciencias Astron{\'o}micas y
  Geof{\'i}sicas,
  Universidad Nacional de La Plata,
  La Plata, Argentina; CONICET, 
  Buenos Aires, Argentina; contrera@fisica.unlp.edu.ar}
\corres{Correspondence: fweber@sdsu.edu}

\abstract{In this work we investigate the effect a crystalline
  quark-hadron mixed phase can have on the neutrino emissivity from
  the cores of neutron stars.  To this end we use relativistic
  mean-field equations of state to model hadronic matter and a
  nonlocal extension of the three-flavor Nambu-Jona-Lasinio model for
  quark matter.  Next we determine the extent of the quark-hadron
  mixed phase and its crystalline structure using the Glendenning
  construction, allowing for the formation of spherical blob, rod, and
  slab rare phase geometries. Finally we calculate the neutrino
  emissivity due to electron-lattice interactions utilizing the
  formalism developed for the analogous process in neutron star
  crusts.  We find that the contribution to the neutrino emissivity
  due to the presence of a crystalline quark-hadron mixed phase is
  substantial compared to other mechanisms at fairly low temperatures
  ($\lesssim 10^9$ K) and quark fractions ($\lesssim 30\%$), and that
  contributions due to lattice vibrations are insignificant compared
  to static-lattice contributions.  There are a number of open issues
  that need to be addressed in a future study on the neutrino emission
  rates caused by electron-quark blob bremsstrahlung. Chiefly among
  them are the role of collective oscillations of matter, electron
  band structures, and of gaps at the boundaries of the Brillouin
  zones on bremsstrahlung, as discussed in the summary section of this
  paper. We hope this paper will stimulate studies addressing these
  issues.}


\begin{document}

\section{Introduction}\label{sec:introduction}

\begin{figure}[htb]
\centering
\includegraphics[scale=0.30]{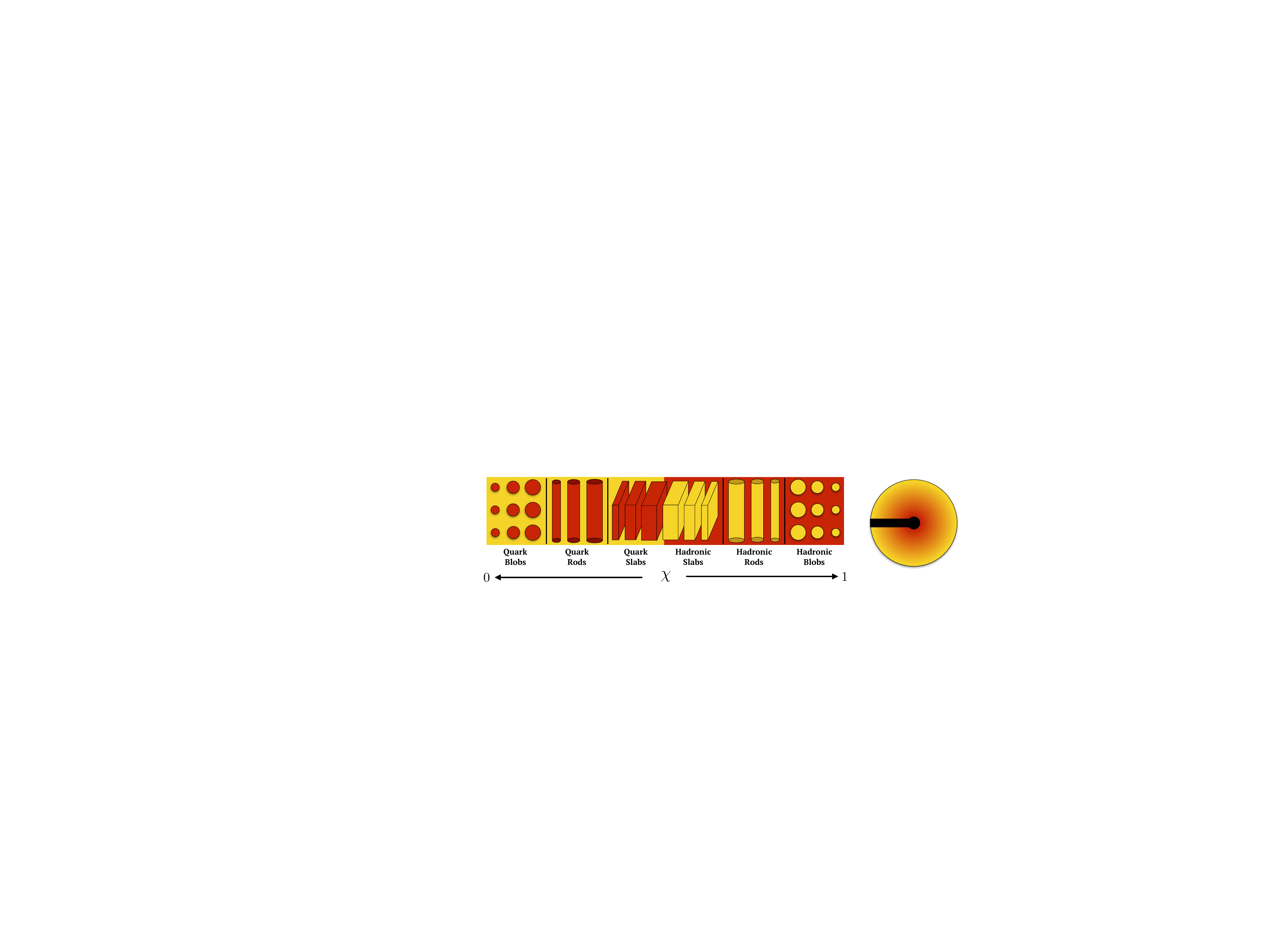}
\caption{Schematic illustrating the rare phase structures that may
  form in the quark-hadron mixed phase 
 \cite{Spinella2016,Spinella2017:thesis}). An increase in the volume
  fraction of quark matter, described by $\chi$, is accompanied by an
  increase in baryon number density and depth within a neutron star.}
\label{fig:mixed-phase-geometries}
\end{figure}

It was shown by Glendenning \cite{Glendenning1992,Glendenning2001}
that if electric charge neutrality in a neutron star
\cite{Page2006,Becker2009,Buballa2014} is treated globally rather than locally, the
possible first order phase transition from hadronic matter to quark
matter in the neutron star core will result in a mixed phase in which
both phases of matter coexist. To minimize the total isospin asymmetry
energy the two phases will segregate themselves, which results in
positively charged regions of hadronic matter and negatively charged
regions of quark matter, with the rare phase occupying sites on a
Coulomb lattice. The situation is schematically illustrated in
Fig.\ \ref{fig:mixed-phase-geometries}.
Further, the competition between the Coulomb and surface energy
densities will cause the matter to arrange itself into energy
minimizing geometric configurations
\cite{Glendenning1992,Glendenning2001}.

The presence of the Coulomb lattice and the nature of the geometric
configurations of matter in the quark-hadron mixed phase may have a
significant effect on the neutrino emissivity from the core. More 
specifically, neutrino-antineutrino pairs will be created by the 
scattering of electrons from these charged lattice structures, 
\begin{equation} \label{eq:electron-structure-bremsstrahlung}
  {\rm e}^- + (Z,A) \rightarrow  {\rm e}^- + (Z,A) + \nu + \bar\nu\,,
\end{equation}
and this will increase the emissivity in the mixed phase.  This
process is analogous to neutrino-pair bremsstrahlung of electrons in
the neutron star crust, where ions exist on a lattice immersed in an
electron gas, and for which there exists a large body of work (see,
for example
\cite{Kaminker1999,Flowers1973,Itoh1983a,Itoh1983b,Itoh1984c,Itoh1984d,Pethick1997}). The
situation is more complicated in the quark-hadron mixed phase, but the
operative interaction is still the Coulomb interaction. Thus, to
estimate the neutrino-pair Bremsstrahlung of electrons from rare phase
structures in the quark-hadron mixed phase we rely heavily on this
body of work (particularly \cite{Kaminker1999}).  We will refer to
this additional mechanism as mixed phase Bremsstrahlung (MPB).

Neutrino emissivity due to the interaction of electrons with a
crystalline quark-hadron mixed phase has been previously studied in
this manner in \cite{Na2012,Spinella2016}. In the present work we use
a set of nuclear equations of state which are in better agreement with the
latest nuclear matter constraints at saturation density than those
utilized in \cite{Spinella2016}, and are consistent with the $2.01\,\mathrm{M}_{\odot}$
mass constraint set by PSR J0348+0432 \cite{Antoniadis2013}. 
To describe quark matter we use the nonlocal SU(3) Nambu-Jona-Lasinio
(n3NJL) model discussed in \cite{Scarpettini2004,Contrera2008,Contrera2010,
Orsaria2013,Orsaria2014,Spinella2016}. The n3NJL parametrization used 
is given as "Set I" in \cite{Sandoval2016}, and is in better agreement
with the empirical quark masses than the parametrization utilized in
\cite{Spinella2016}.  We consider three geometries for
the range of possible structures in the mixed phase including spherical blobs,
rods, and slabs, and calculate the associated static lattice
contributions to the neutrino emissivity. Phonon contributions to the
emissivity for rod and slab geometries are not considered, though a
comparison of the phonon and static lattice contributions for
spherical blobs is given and indicates that phonon contributions may
not be significant. Finally, the extent of the conversion to quark
matter in the core was determined in \cite{Spinella2017:thesis}, and this
allows for a comparison between emissivity contributions from standard
and enhanced neutrino emission mechanisms including the direct Urca (DU),
modified Urca (MU), and baryon-baryon and quark-quark Bremsstrahlung
(NPB) processes, and contributions from electron-lattice interactions.
For a detailed summary including the equations and coefficients
used for the calculation of the standard and enhanced neutrino
emission mechanisms see \cite{Spinella2017:thesis}.

The results for different parametrizations are numerous and qualitatively
similar, so the DD2 parametrization will be presented exclusively
in this paper. The results of the other parametrizations can be found in
\cite{Spinella2017:thesis}.

\section{Improved set of Models for the Nuclear Equation of State}\label{sec:eos}

Hadronic matter is modeled in the framework of the relativistic
nonlinear mean-field (RMF) approach \cite{Boguta1977,Boguta1983},
which describes baryons interacting through the exchange of scalar,
vector, and isovector mesons (for details, see
\cite{Spinella2017:thesis,Spinella2016,Mellinger2017}).  The RMF
approach is parametrized to reproduce the following properties of symmetric
nuclear matter at saturation density $n_0$ (see Table
\ref{table:parametrizations}): the binding energy per
nucleon ($E_0$), the nuclear incompressibility ($K_0$), the isospin
asymmetry energy ($J$), and the effective mass ($m^*/m_N$).  In
\begin{table}[htb]
\begin{center}
\caption{Properties of symmetric nuclear matter at saturation density for the
  hadronic parametrizations of this work.}
\begin{tabular}{ccccc}
\toprule $~~$\textbf{Saturation Property}$~~$ & $~~$\textbf{SWL}
\cite{Spinella2017:thesis} $~~$ & $~~$\textbf{GM1L}
\cite{Spinella2017:thesis,Glendenning1992}$~~$ & \textbf{DD2}
\cite{Typel2010}$~~$ &\textbf{ME2} \cite{Lalazissis2005}$~~$ \\
\noalign{\smallskip}\hline\noalign{\smallskip}
$n_0$ (fm$^{-3}$) & 0.150 & 0.153 & 0.149 & 0.152 \\
$E_0$ (MeV) & $-16.00$ & $-16.30$ & $-16.02$ & $-16.14$ \\
$K_0$ (MeV) & 260.0 & 300.0 & 242.7 & 250.9 \\
$m^*/m_N$ & 0.70 & 0.70 & 0.56 & 0.57 \\
$J$ (MeV) & 31.0 & 32.5 & 32.8 & 32.3 \\
$L_0$ (MeV) & 55.0 & 55.0 & 55.3 & 51.3 \\
\bottomrule
\end{tabular}
\label{table:parametrizations}
\end{center}
\end{table}
addition, the RMF parametrizations used in this work employ a
density-dependent isovector-meson-baryon coupling constant that can be
fit to the slope of the asymmetry energy ($L_0$) at $n_0$. The scalar-
and vector-meson-baryon coupling constants of the density-dependent
relativistic mean-field models DD2 and ME2 are fit to properties of
finite nuclei \cite{Spinella2017:thesis,Typel2010,Lalazissis2005}.
These models are an extension of the standard RMF approach
that account for medium effects by making the meson-baryon
coupling constants dependent on the local baryon number density
\cite{Fuchs1995}.
The density-dependence of the meson-baryon coupling constants is given
by \begin{equation} \label{eq:ddrmf-couplings} g_{iB}(n) =
  g_{iB}(n_0)f_i(x),
\end{equation}
where $i\,\in\,\{\sigma,\omega,\rho\}$, $x=n/n_0$, and $f_{i}(x)$
provides the functional form for the density dependence. The most
commonly utilized ansatz for $f_{i}(x)$ are given by \cite{Typel1999}
\begin{equation} \label{eq:coupling-density-dependence}
	f_i(x) = a_i \frac{1+b_i(x+d_i)^2}{1+c_i(x+d_i)^2}\,,
\end{equation}
for $i\in\{\sigma,\omega\}$, and
\begin{equation} \label{eq:coupling-density-dependence-rho}
	f_{\rho}(x) = \mathrm{exp}\left[-a_{\rho}\left(x-1\right)\right] \, .
\end{equation}
The nine parameters of the density dependence
($a_{\sigma},b_{\sigma},c_{\sigma},
d_{\sigma},a_{\omega},b_{\omega},c_{\omega},d_{\omega},a_\rho$), the
values of the meson-nucleon couplings at $n_0$ ($g_{\sigma
  N}(n_0),g_{\omega N}(n_0), g_{\rho B}(n_0)$), and the mass of the
scalar meson ($m_{\sigma}$) are all fit to properties of symmetric
nuclear matter at $n_0$ and to the properties of finite nuclei
including but not limited to binding energies, charge and diffraction
radii, spin-orbit splittings, and neutron skin thickness (see
\cite{Lalazissis2005,Typel2005}).

In addition to the nucleons, hyperons and delta isobars ($\Delta$s) are
also considered in the composition of hadronic matter. The scalar-meson-hyperon
coupling constants are fit to the following hypernuclear potentials at
saturation (see \cite{Spinella2017:thesis} and references therein),
\begin{equation} \label{eq:scalar-hyperon-couplings}
  U_{\Lambda}^{(N)} = -28\,\mathrm{MeV},\,\,U_{\Sigma}^{(N)} =
  +30\,\mathrm{MeV},\,\, U_{\Xi}^{(N)} = -18\,\mathrm{MeV}\,.
\end{equation}
The vector-meson-hyperon coupling constants are taken to be those
given by the ESC08 model in SU(3) symmetry
\cite{Spinella2017:thesis,Rijken2010,Miyatsu2013},
\begin{equation} \label{eq:su3}
  g_{\omega\Lambda} = g_{\omega\Sigma} \approx 0.79\,g_{\omega N},\,\,\,
  g_{\omega\Xi} \approx 0.59\,g_{\omega N}\,.
\end{equation}
The scalar- and vector-meson-$\Delta$ coupling constants are given as follows,
\begin{equation} \label{eq:delta-couplings}
  x_{\sigma\Delta}=x_{\omega\Delta}=1.1,\,x_{\rho\Delta} = 1.0\,.
\end{equation}
Finally, the isovector-meson-hyperon and isovector-meson-$\Delta$ coupling
constants are taken to be universal, with the differences in the baryon isospin
accounted for by the isospin operator in the lagrangian.

\section{Crystalline Structure of the Quark-Hadron Mixed Phase}\label{sec:structure}

A mixed phase of hadronic and quark matter will arrange itself so as
to minimize the total energy of the phase. Under the condition of
global charge neutrality this is the same as minimizing the
contributions to the total energy due to phase segregation, which
includes the surface and Coulomb energy contributions.  Expressions
for the Coulomb ($\epsilon_C$) and surface ($\epsilon_S$) energy
densities can be written as \cite{Glendenning1992,Glendenning2001}
\begin{eqnarray}
  \mathcal{E}_C &=& 2\pi e^2 \left[ q_H(\chi) - q_Q(\chi)
    \right]^2 r^2 x f_D(x) \, ,
    \label{eq:eps_c} \\
    \mathcal{E}_S &=& D x \alpha(\chi)/r\, ,
    \label{eq:eps_s}
\end{eqnarray}
where $q_H$ ($q_Q$) is the hadronic (quark) phase charge density, $r$
is the radius of the rare phase structure, and $\alpha(\chi)$ is the
surface tension between the two phases.  The parameter $\chi$, which
varies between 0 and 1, represents the volume fraction of quark matter
at a given density.  The quantities $x$ and $f_D(x)$ in
\eqref{eq:eps_c} are defined as
\begin{equation}
  x = \mathrm{min}(\chi,1-\chi) 
\end{equation}
and
\begin{equation}
  f_D(x) = \frac{1}{D+2} \left[ \frac{1}{D-2} (2-D\, x^{1-2/D}) + x
    \right]\,,
  \end{equation}
where $D$ is the dimensionality of the lattice.  The phase
rearrangement process will result in the formation of geometrical
structures of the rare phase distributed in a crystalline lattice that
is immersed in the dominant phase (see
Fig.\ \ref{fig:mixed-phase-geometries}).  The rare phase structures
are approximated for convenience as spherical blobs, rods, and slabs
\cite{Glendenning1992,Glendenning2001}. The spherical blobs occupy
sites in a three dimensional ($D=3$) body centered cubic (BCC)
lattice, the rods in a two dimensional ($D=2$) triangular lattice, and
the slabs in a simple one dimensional ($D=1$) lattice
\cite{Kaminker1999}. At $\chi = 0.5$ both hadronic and quark matter
exist as slabs in the same proportion, and at $\chi > 0.5$ the
hadronic phase becomes the rare phase with its geometry evolving in
reverse order (from slabs to rods to blobs).

Direct determination of the surface tension of the quark-hadron
interface is problematic because of difficulties in constructing a
single theory that can accurately describe both hadronic matter and
quark matter.  Therefore, we employ an approximation proposed by Gibbs
where the surface tension is taken to be proportional to the
difference in the energy densities of the interacting phases
\cite{Glendenning1992,Glendenning2001},
\begin{equation}
  \alpha(\chi)=\eta L \left[\mathcal{E}_Q(\chi)-\mathcal{E}_H(\chi)\right]\, ,
\end{equation}
where $L$ is proportional to the surface thickness which should be on
the order of the range of the strong interaction (1 fm), and $\eta$ is
a proportionality constant. In this work we maintain the energy
density proportionality but set the parameter $\eta = 0.08$ so that
the surface tension falls below 70 MeV fm$^{-2}$
for all parametrizations, a reasonable upper limit for the existence of a
\begin{figure}[tb]
\centering
\includegraphics[scale=0.65]{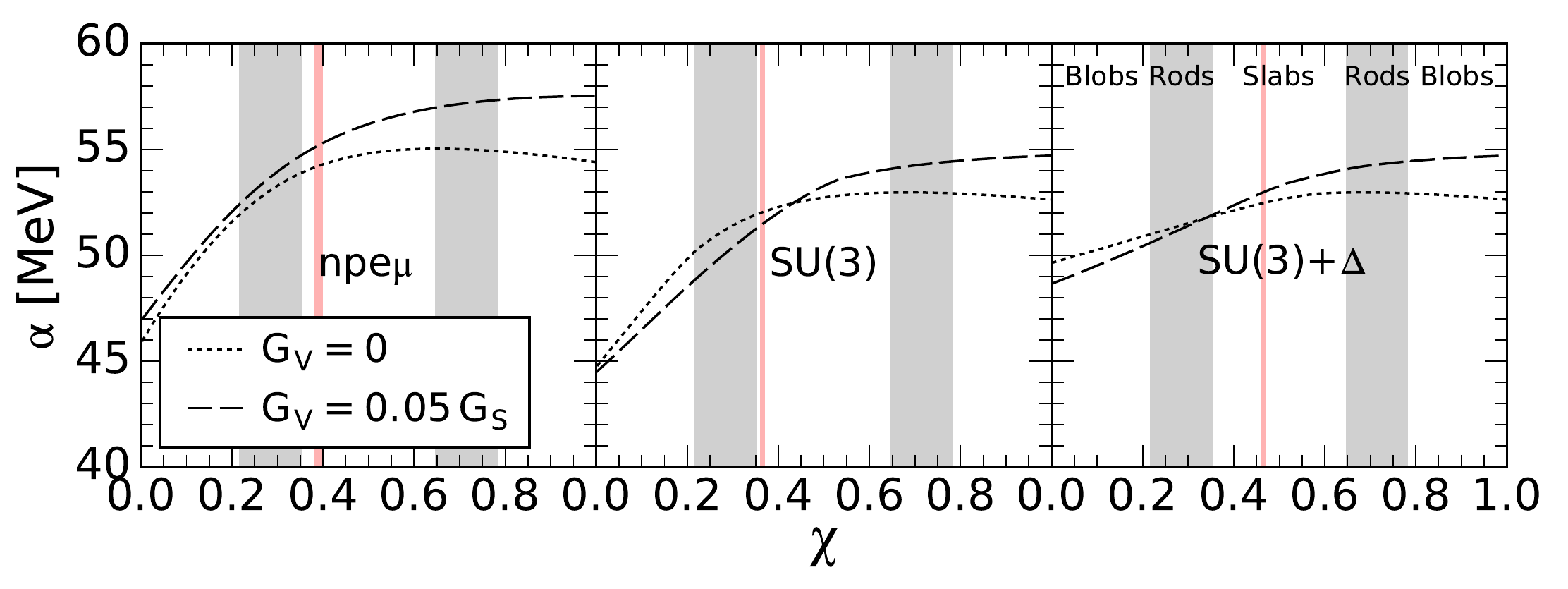}
\caption{Surface tension $\alpha$ in the quark-hadron mixed phase for
  the DD2 parametrization \cite{Spinella2017:thesis}.  The red
  shading indicates the range for the maximum quark fraction
  $\chi_{\mathrm{max}}$ for the two values of the quark vector
  coupling constant $G_V$. (Left panel) Only nucleons and leptons are
  included in the hadronic phase. (Center panel) Hyperons are included
  in the hadronic phase. (Right panel) Delta isobars are included in
  addition to hyperons in the hadronic phase.
  Similar figures for the SWL, GM1L, and ME2 parametrizations can be
    found in Ref.\ \cite{Spinella2017:thesis}.}
\label{fig:surface-tension}
\end{figure}
quark-hadron mixed phase \cite{Yasutake2014}. The surface tension as a function
of $\chi$ is given in Figure \ref{fig:surface-tension} for the nuclear DD2
parametrization, introduced in Sect.\ \ref{sec:eos}.

We note that, in this work, we restricted ourselves to considering
$G_V$ values that are in the range of $0 < G_V < 0.05 G_S$, as this
choice leads to gravitational masses of neutron stars with
quark-hybrid compositions that satisfy the $2_\odot$
constraint. Exploring the possibility of larger $G_v$ values would
certainly be worthwhile, but this is beyond the scope of this work.

The size of the rare phase structures is given by the radius ($r$) and
is determined by minimizing the sum of the Coulomb and surface
energies, $\partial(\mathcal{E}_C+\mathcal{E}_S) / \partial r$, and
solving for $r$ \cite{Glendenning1992,Glendenning2001},
\begin{equation}
  r(\chi) = \left(\frac{D\alpha(\chi)}
    {4\pi e^2f_D(\chi)\left[q_H(\chi)
  - q_Q(\chi)\right]^2}\right)^\frac{1}{3}.
\end{equation}
Rare phase structures are centered in the primitive cell of the lattice,
taken to be a Wigner-Seitz cell of the same geometry as the
rare phase structure. The Wigner-Seitz cell radius $R$ is set so that the 
primitive cell is charge neutral,
\begin{equation} \label{eq:wigner-seitz-radius}
  R(\chi) = rx^{-1/D} \, .
\end{equation}

\begin{figure}[tb]
\centering
\includegraphics[scale=0.65]{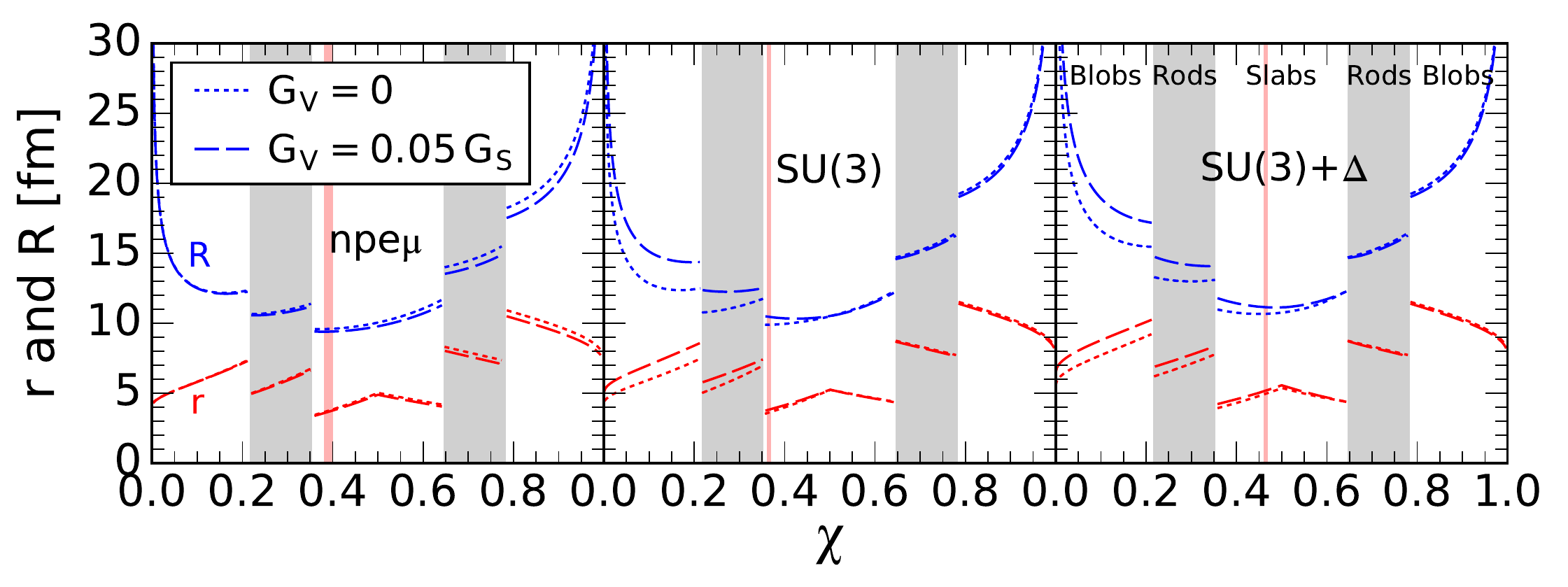}
\caption{Radius of the rare phase structure $r$ and Wigner-Seitz cell
  $R$ in the quark-hadron mixed phase for the DD2 parametrization
  \cite{Spinella2017:thesis}.  See Figure \ref{fig:surface-tension}
  for additional details.  Similar figures for the SWL, GM1L, and ME2
  parametrizations can be found in Ref.\ \cite{Spinella2017:thesis}.}
\label{fig:structure-radii}
\end{figure}

Figure \ref{fig:structure-radii} shows $r$ and $R$ as a function of the quark
fraction in the mixed phase.  Both $r$ and $R$ 
increase with an increase in the baryonic degrees of freedom, particularly when
$\chi \lesssim 0.5$ and the vector interaction is included.
Note that the blob radius should vanish for $\chi \in \{0,1\}$, but does not due
to the approximate nature of the geometry function $f_D(\chi)$ \cite{Na2012}.
The number density of rare phase blobs will be important for calculating
the phonon contribution to the emissivity. Since there is one rare phase
blob per Wigner-Seitz cell, the number density of rare phase blobs ($n_b$)
is simply the reciprocal of the Wigner-Seitz cell volume,
\begin{equation} \label{eq:blobdensity}
  n_b = (4\pi R^3/3)^{-1} \, .
\end{equation}

\begin{figure}[tb]
\centering
\includegraphics[scale=0.65]{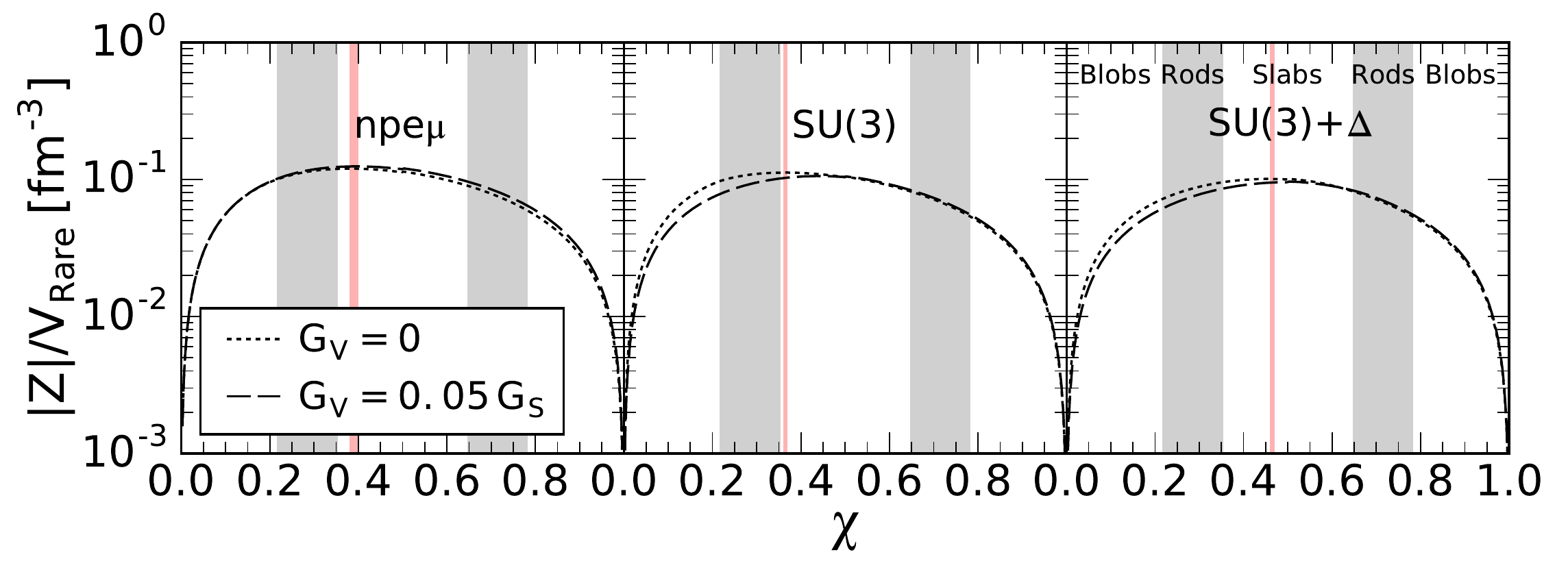}
\caption{Charge number per unit volume of the rare phase structures
  for the DD2 parametrization \cite{Spinella2017:thesis}.  See Figure
  \ref{fig:surface-tension} for additional details.  Similar figures
  for the SWL, GM1L, and ME2 parametrizations can be found in
  Ref.\ \cite{Spinella2017:thesis}.}
\label{fig:structure-charge}
\end{figure}

The density of electrons in the mixed phase is taken to be uniform
throughout. Charge densities in both the rare and dominant phases
are also taken to be uniform, an approximation supported by a recent study
by Yasutake {\it et al.}\ \cite{Yasutake2014}. The uniformity of charge
in the rare phase also justifies the use of the nuclear form factor
($F(q)$) presented in Section \ref{sec:emissivity}.
The total charge number per unit volume ($\left|Z\right|/V_{\mathrm{Rare}}$)
of the rare phase structures is given in Figure \ref{fig:structure-charge}.

\section{Neutrino Emissivity due to a Crystalline Quark-Hadron Lattice}
\label{sec:emissivity}

We begin this section with a brief discussion of the neutrino
emissivity due to a crystalline quark-hadron lattice
\cite{Spinella2016}.  Modeling the complex interactions of electrons
with a background of neutrons, protons, hyperons, muons, and quarks is
an exceptionally complicated problem. However, to make a determination
of the neutrino emissivity that is due to electron-lattice
interactions in the quark-hadron mixed phase we need only consider the
Coulomb interaction between them.  This simplifies the problem
greatly, as a significant body of work exists for the analogous
process of electron-ion scattering that takes place in the crusts of
neutron stars.

To determine the state of the lattice in the quark-hadron mixed
phase we use the dimensionless ion coupling parameter given by
\begin{equation} \label{eq:ion-coupling-parameter}
  \Gamma = \frac{Z^2e^2}{R k_B T}\, .
\end{equation}
Below $\Gamma_{\mathrm{melt}} = 175$ the lattice behaves as a Coulomb
liquid, and above as a Coulomb crystal \cite{Ogata1987,HaenselBook}.
It was shown in Ref.\ \cite{Na2012} that the emissivity due to
electron-blob interactions in the mixed phase was insignificant
compared to other contributions at temperatures above $T \gtrsim
10^{10}$ K. Therefore, in this work we consider temperatures in the
range $10^7\,\mathrm{K} \le T \le 10^{10}\,\mathrm{K}$. At these
temperatures the value of the ion coupling parameter is well above
$\Gamma_{\mathrm{melt}}$, and so the lattice in the quark-hadron mixed
phase is taken to be a Coulomb crystal.

To account for the fact that the elasticity of scattering events is
temperature dependent we need to compute the Debye-Waller factor,
which is known for spherical blobs only and requires the plasma
frequency and temperature given by
\begin{equation}
  \omega_p = \sqrt{\frac{4\pi Z^2e^2n_b}{m_b}}\, ,
\end{equation}
\begin{equation}
  T_p = \frac{\hbar \omega_p}{k_B}\, ,
\end{equation}
where $m_b$ is the mass of a spherical blob \cite{Kaminker1999}.
The Debye-Waller factor is then given by 
\begin{equation} \label{eq:debye}
  W(q) = \begin{cases}
    \frac{a q^2}{8k_e^2}\! \left(1.399\, {\rm e}^{-9.1 t_p} + 12.972\,t_p\right)
    & \mathrm{spherical~blobs}\, , \\
    0 & \mathrm{rods~and~slabs}\, ,
  \end{cases}
\end{equation}
where $q=|\boldsymbol{q}|$ is a phonon or scattering wave vector, $a =
4\hbar^2 k_e^2 /(k_B T_p m_b)$, and $t_p = T/T_p$
\cite{Kaminker1999,Baiko1995}. In order to smooth out the charge
distribution over the radial extent of the rare phase structure we
adopt the nuclear form factor given in \cite{Kaminker1999},
\begin{equation} \label{eq:formfactor}
  F(q) =
  \frac{3}{(qR^3)}\left[\mathrm{sin}(qR)-qR\,\mathrm{cos}(qR)\right]
  \, .
\end{equation}
Screening of the Coulomb potential by electrons is taken into account
by the static dielectric factor $\epsilon(q,0)=\epsilon(q)$, given in
Ref.\ \cite{Itoh1983a}.  However, the charge number of the rare phase
structures is high and the electron number density is low, so setting
this factor to unity has no noticeable effect on the calculated
neutrino emissivity. Finally, the effective interaction is given by
\cite{Kaminker1999}
\begin{equation} \label{eq:potential}
  V(q) = \frac{4\pi e \rho_Z F(q)}{q^2 \epsilon(q,0)} \, {\rm
    e}^{-W(q)} \, .
\end{equation}


General expressions for the neutrino emissivity due to the MPB electron-lattice
interactions were derived by Haensel {\it et al.}\ \cite{Haensel1996} for spherical
blobs and by Pethick {\it et al.}\ \cite{Pethick1997} for rods and slabs,
\begin{equation}
  \epsilon_{\mathrm{MPB}}^{\text{blobs}} \approx 5.37\times10^{20}\,nT_9^6Z^2L
  \;\;\text{erg\,s}^{-1}\text{\,cm}^{-3}\,,
\end{equation}
\begin{equation}
  \epsilon_{\mathrm{MPB}}^{\text{rods,slabs}} \approx 4.81\times10^{17}\;k_eT_9^8J
  \;\;\text{erg\,s}^{-1}\text{\,cm}^{-3}\, ,
    \label{eq:23}
\end{equation}
where $L$ and $J$ are dimensionless quantities that scale the
emissivities. Both $L$ and $J$ contain a contribution due to the
static lattice (Bragg scattering), but we consider the additional
contribution from lattice vibrations (phonons) for spherical blobs, so
$L = L_{\mathrm{sl}} + L_{\mathrm{ph}}$.  We note that the $T^8$
temperature dependence in Eq.\ (\ref{eq:23}) is somewhat deceiving
since the $J$ factor also depends on temperature and, for a wide range
of parameters, is proportional to $1/T^2$. In effect, the neutrino
emissivity $\epsilon_{\mathrm{MPB}}^{\text{rods,slabs}}$ is therefore
proportional to $T^6$.

\subsection{Phonon Contribution to Neutrino Emissivity}

\begin{figure}[tb]
\centering
\includegraphics[scale=0.65]{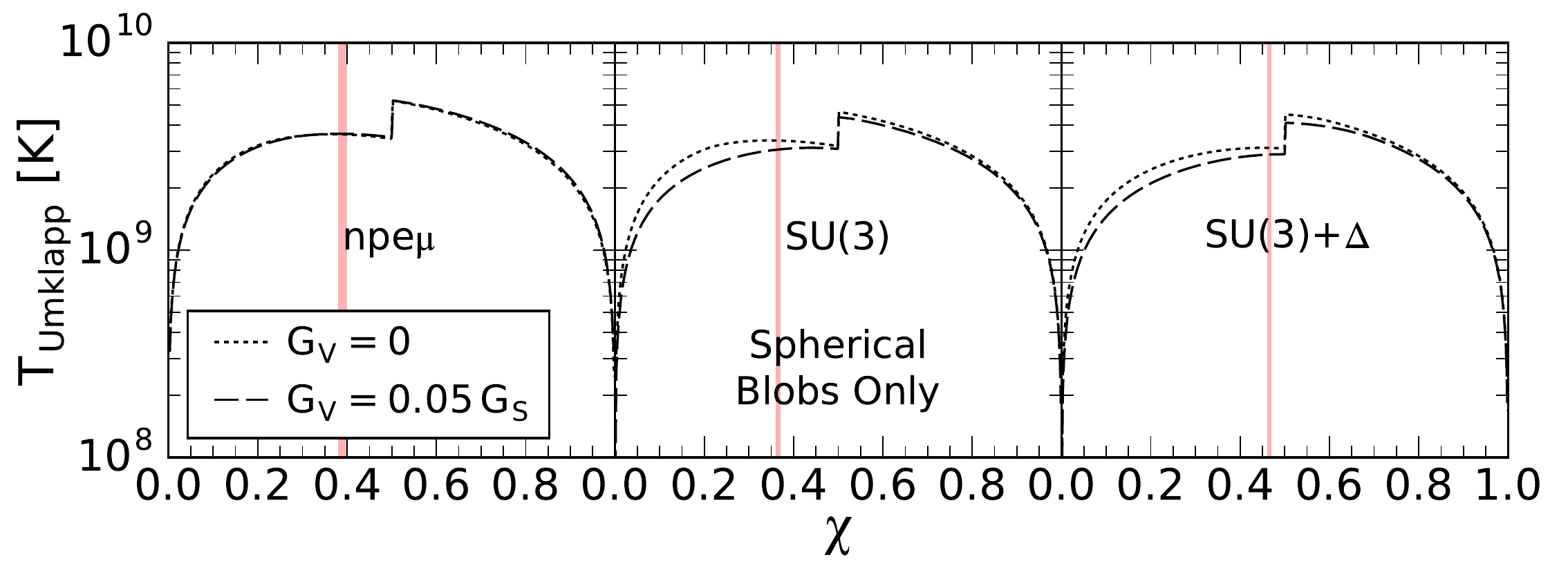}
\caption{Temperature below which Umklapp processes are frozen out
  ($T_{\mathrm{Umklapp}}$), and contributions to the neutrino
  emissivity due to electron-phonon interactions become negligible for
  the DD2 parametrization \cite{Spinella2017:thesis}.  See Figure
  \ref{fig:surface-tension} for additional details.  Similar figures
  for the SWL, GM1L, and ME2 parametrizations can be found in
  Ref.\ \cite{Spinella2017:thesis}.}
\label{fig:umklapp}
\end{figure}

The expressions for determining the neutrino emissivity due to
interactions between electrons and lattice vibrations (phonons) in a
Coulomb crystal, with proper treatment of multi-phonon processes, were
obtained by Baiko {\it et al.}\ \cite{Baiko1998} and simplified by Kaminker
{\it et al.}\ \cite{Kaminker1999}. The phonon contribution to the
emissivity is primarily due to Umklapp processes in which a phonon is
created (or absorbed) by an electron that is simultaneously Bragg
reflected, resulting in a scattering vector $\boldsymbol{q}$ that lies
outside the first Brillouin zone, $q_0 \gtrsim (6\pi^2n_b)^{1/3}$
\cite{Raikh1982,ZimanBook}, where $n_b$ is given by
Eq.\ \eqref{eq:blobdensity}.

The contribution to MPB due to phonons is
contained in $L_{\mathrm{ph}}$ and given by Eq.\ (21) in
Ref.\ \cite{Kaminker1999},
\begin{equation} \label{eq:lph}
  L_{\mathrm{ph}} = \int_{y_0}^1 dy \frac{S_{\mathrm{eff}}(q)
  |F(q)|^2}{y|\epsilon(q,0)|^2}\left(1+\frac{2y^2}{1-y^2}
  \mathrm{ln}\,y\right)\, ,
\end{equation}
where $y=q/(2k_e)$, and the lower integration limit $y_0$ excludes
momentum transfers inside the first Brillouin zone.  The structure
factor $S_{\mathrm{eff}}$ is given by (24) and (25) in
Ref.\ \cite{Kaminker1999}),
\begin{equation} \label{eq:seff}
  S_{\mathrm{eff}}(q) = 189\left(\frac{2}{\pi}\right)^5 {\rm e}^{-2W}
  \int_0^{\infty} d\xi \, \frac{1-40\xi^2+80\xi^4}
  {\left(1+4\xi^2\right)^5
    \mathrm{cosh}^2\left(\pi\xi\right)}
  \times \left( {\rm e}^{\Phi(\xi)}-1\right)\, ,
\end{equation}
\begin{equation} \label{eq:phi}
  \Phi(\xi) = \frac{\hbar q^2}{2m_b}
  \left\langle \frac{\mathrm{cos}\left(\omega_st\right)}
  {\omega_s \mathrm{sinh}\left(\hbar \omega_s/2k_BT\right)} \right\rangle\, ,
\end{equation}
where $\xi = tk_BT/\hbar$ and $\langle \ldots \rangle$ denotes
averaging over phonon frequencies and modes,
\begin{equation} \label{eq:fs}
  \langle f_s(\boldsymbol{k}) \rangle = \frac{1}{3V_B}\sum\limits_s
  \int_{V_B} d\boldsymbol{k}\, f_s(\boldsymbol{k})\, .
\end{equation}
It is assumed that there are three phonon modes $s$, two linear
transverse and one longitudinal. The frequencies of the transverse
modes are given by $\omega_{t_i} = a_ik$, where $i=1,2$, $a_1 =
0.58273$, and $a_2 = 0.32296$.  The frequency of the longitudinal mode
$\omega_l$ is determined by Kohn's sum rule, $\omega_l^2 = \omega_p^2
- \omega_{t_1}^2 - \omega_{t_2}^2$ \cite{Mochkovitch1979}.

Umklapp processes proceed as long as the temperature 
$T_{\mathrm{Umklapp}} \gtrsim T_pZ^{1/3} e^2/(\hbar c)$, below which electrons 
can no longer be treated in the free electron approximation \cite{Raikh1982}.
This limits the phonon contribution to the neutrino emissivity
to only a very small range in temperature for a crystalline quark-hadron
mixed phase (see Figure \ref{fig:umklapp}), and renders it negligible compared to
the static lattice contribution as will be shown in the next section.

\subsection{Static Lattice Contribution to Neutrino Emissivity}

Pethick and Thorsson \cite{Pethick1997} found that with proper
handling of electron band-structure effects the static lattice
contribution to the neutrino emissivity in a Coulomb crystal was
significantly reduced compared to calculations performed in the free
electron approximation. Kaminker {\it et al.}\ \cite{Kaminker1999} presented
simplified expressions for calculating the static lattice contribution
($L_{\mathrm{sl}}$) using the formalism developed in
Ref.\ \cite{Pethick1997}.  The dimensionless quantities
$L_{\mathrm{sl}}$ and $J$ that scale the neutrino emissivities for
spherical blobs and rods/slabs, respectively, are given by
\begin{equation} \label{eq:lsl}
  L_{\mathrm{sl}} = \frac{1}{12Z} \sum_{K \ne 0}
  \frac{(1-y_K^2)}{y_K^2}\frac{|F(K)|^2}{|\epsilon(K)|^2}\,I(y_K,t_V)
  \, {\rm e}^{-2W(K)} 
\end{equation}
and
\begin{equation} \label{eq:j}
  J = \sum_{K \ne 0}\frac{y_K^2}{t^2_V}I(y_K,t_V)\, ,
\end{equation}
where $K=|\boldsymbol{K}|$ is a scattering vector and restricted to
linear combinations of reciprocal lattice vectors, $y_K=K/(2k_e)$,
$t_V = k_BT/\left[|V(K)|(1-y_K^2)\right]$, and $I(y_K,t_V)$ is given
by Eq.\ (39) in Ref.\ \cite{Kaminker1999}.  The sum over $K$ in
\eqref{eq:lsl} and \eqref{eq:j} terminates when $K > 2k_e$,
prohibiting scattering vectors that lie outside the electron Fermi
surface.

\begin{figure}[tb]
\centering
\includegraphics[scale=0.50]{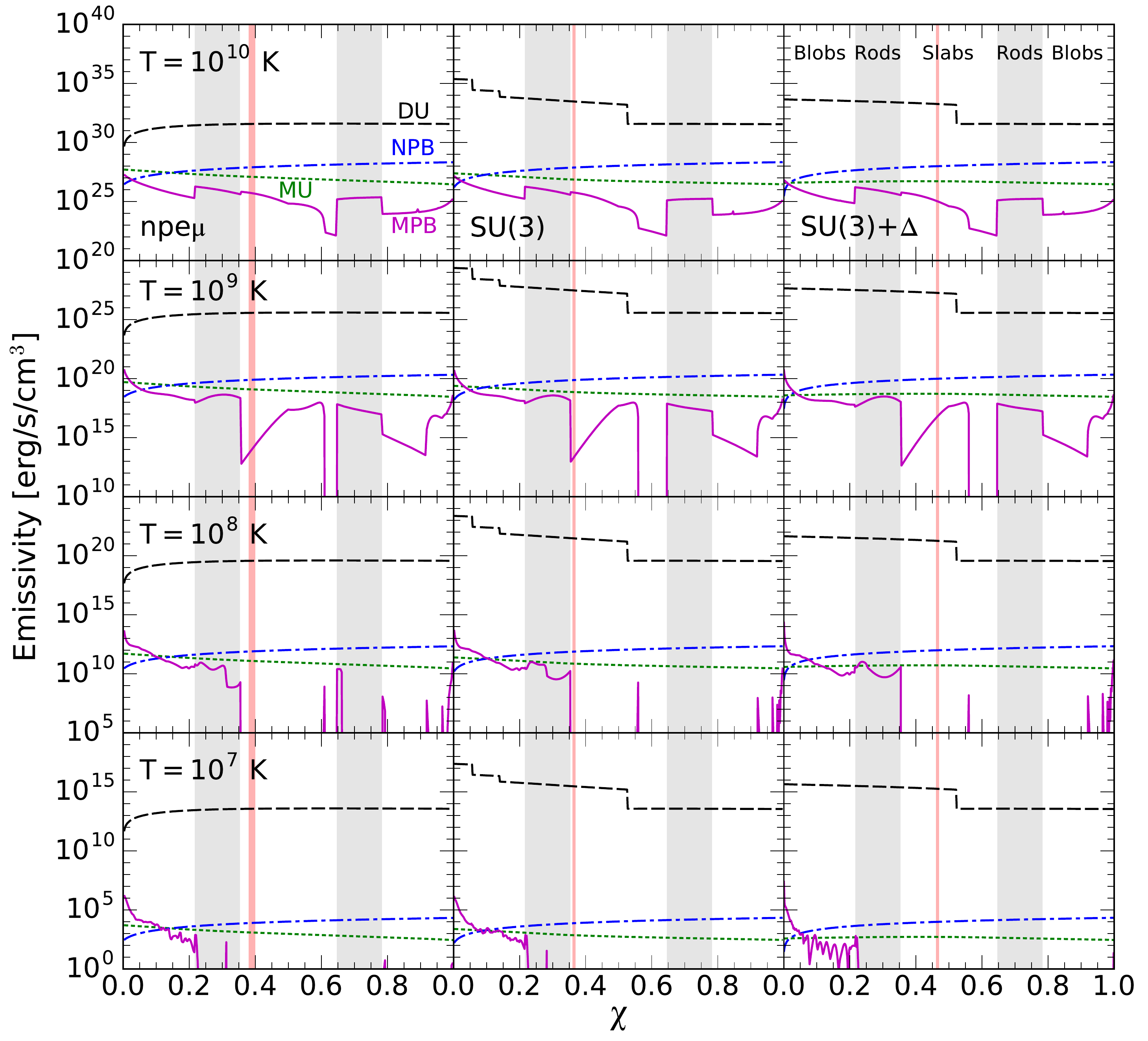}
\caption{Neutrino emissivity in the quark-hadron mixed phase for the
  DD2 parametrization with $G_V = 0$
  \cite{Spinella2017:thesis}. Contributions due to mixed phase
  Bremsstrahlung (MPB), nucleon-nucleon and quark-quark neutrino pair
  Bremsstrahlung (NPB), the nucleon and quark modified Urca processes
  (MU), and the hyperon and quark direct Urca (DU) processes are
  included.  See Figure \ref{fig:surface-tension} for additional
  details.  Similar figures for the SWL, GM1L, and ME2
  parametrizations can be found in Ref.\ \cite{Spinella2017:thesis}.}
\label{fig:emissivity-DD2-gv00}
\end{figure}

\begin{figure}[tb]
\centering
\includegraphics[scale=0.50]{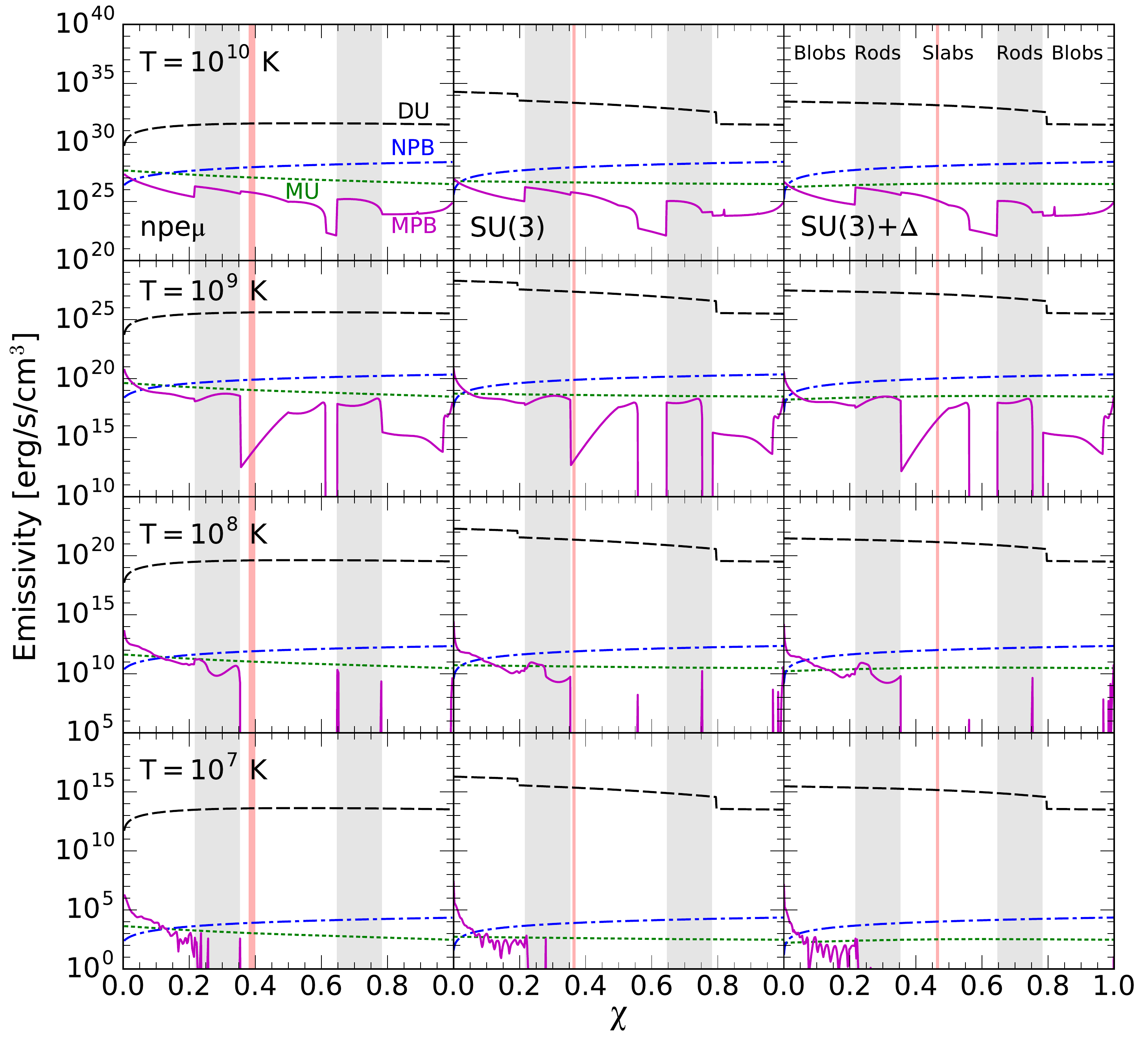}
\caption{Neutrino emissivity in the quark-hadron mixed phase for the
  DD2 parametrization with $G_V = 0.05\,G_S$
  \cite{Spinella2017:thesis}. Contributions due to mixed phase
  Bremsstrahlung (MPB), nucleon-nucleon and quark-quark neutrino pair
  Bremsstrahlung (NPB), the nucleon and quark modified Urca processes
  (MU), and the hyperon and quark direct Urca (DU) processes are
  included.  See Figure \ref{fig:surface-tension} for additional
  details.  Similar figures for the SWL, GM1L, and ME2
  parametrizations can be found in Ref.\ \cite{Spinella2017:thesis}.}
    \label{fig:emissivity-DD2-gv05}
\end{figure}

\begin{figure}[tb]
\centering
\includegraphics[scale=0.65]{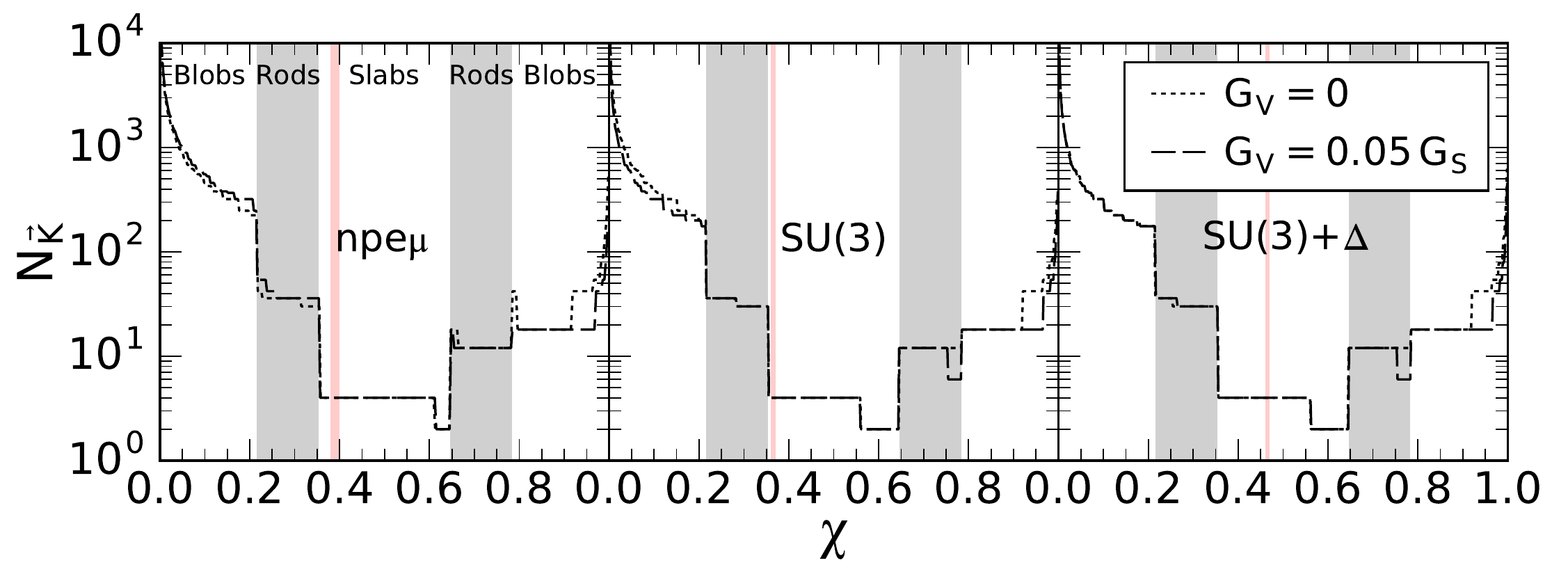}
\caption{The number of scattering vectors that satisfy the condition
  $K<2k_e$ as a function of the quark fraction
  \cite{Spinella2017:thesis} (see Figure \ref{fig:surface-tension} for
  additional details).  Similar figures for the SWL, GM1L, and ME2
  parametrizations can be found in Ref.\ \cite{Spinella2017:thesis}.}
    \label{fig:lattice-vectors}
\end{figure}

\section{Neutrino Emissivity Results}\label{sec:results}

The neutrino emissivities due to MPB and the additional emissivity
mechanisms are given in Figures \ref{fig:emissivity-DD2-gv00} and
\ref{fig:emissivity-DD2-gv05} for $G_V = 0$ and $G_V = 0.05\,G_S$
respectively at temperatures between $10^7 - 10^{10}$ K. The MPB
emissivity is for most of the mixed phase the weakest of the
emissivity mechanisms, peaking at low $\chi$ (at $\chi \lesssim 0.05$
the MPB emissivity may be overestimated due to the limitations of the
dimensionality function), and appears to be slightly larger when
hyperons and $\Delta$s are included in the composition. Including the
vector interaction ($G_V = 0.05\,G_S$) also results in a slight
increase in the MPB emissivity. Both additional baryonic degrees of
freedom and inclusion of the vector interaction delay the onset of the
quark-hadron phase transition, and therefore it may be concluded that
the greater the density in the mixed phase the greater the
contribution to the emissivity from MPB.  The MPB emissivity is most
comparable to the modified Urca emissivity, particularly at
$10^8-10^9$~K.

\begin{figure}[tb]
\centering
\includegraphics[scale=0.65]{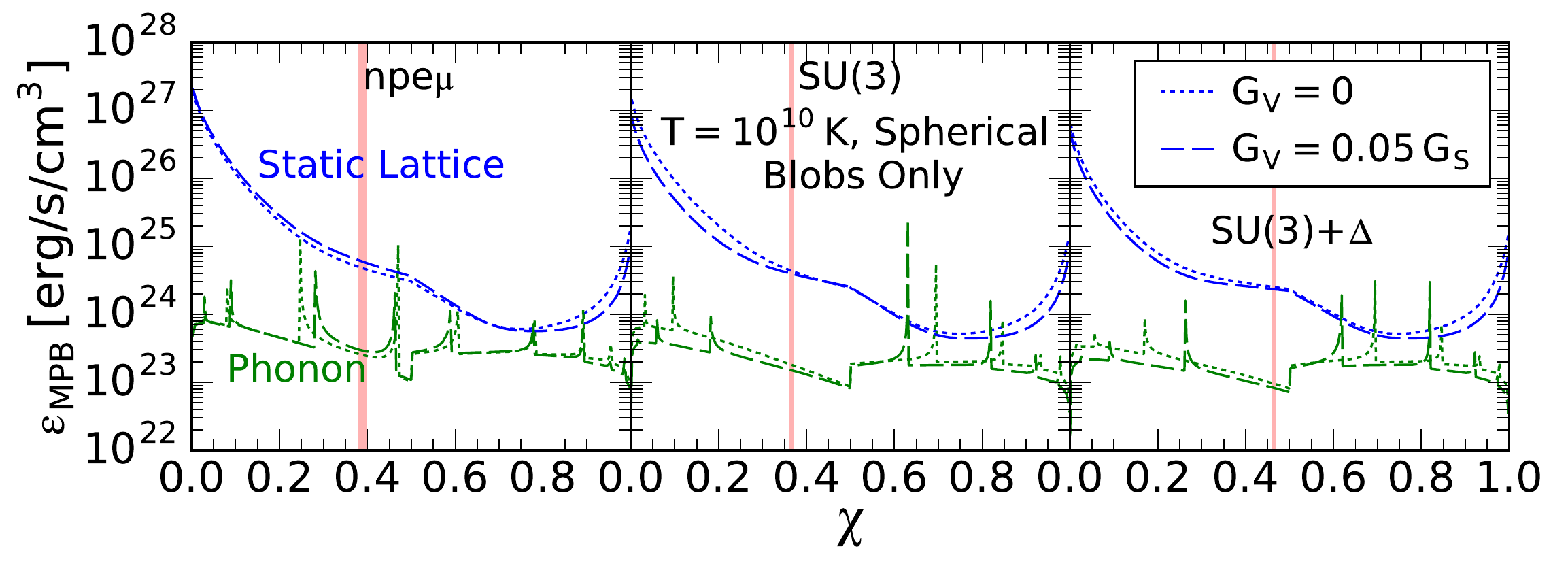}
\caption{Comparison of the static lattice and phonon contributions to
  the neutrino emissivity at $T=10^{10}~{\rm K}$ for the spherical blob
  geometry only and the DD2 parametrization
  \cite{Spinella2017:thesis}.}
    \label{fig:lattice-phonon-comparison}
\end{figure}

Electron-phonon interactions contribute to the MPB emissivity when the
mixed phase consists of spherical blobs ($\chi \lesssim 0.21$ and
$\chi \gtrsim 0.79$) and only when $T > T_{\mathrm{Umklapp}}$ (Figure
\ref{fig:umklapp}), which for the given choices of temperature implies
$T = 10^{10}$~K.  Figure \ref{fig:lattice-phonon-comparison} shows
that the static-lattice contribution to the MPB emissivity dominates
the phonon contribution rendering it negligible, particularly at quark
fractions relevant to the neutron stars of this work ($\chi <
0.5$). Therefore, the MPB emissivity is almost entirely due to the
static-lattice contribution (Bragg scattering).

Equations \eqref{eq:lsl} and \eqref{eq:j} indicate that the 
static-lattice contribution to the MPB emissivity is calculated as a sum over
scattering vectors $\boldsymbol{K}$ that satisfy $K < 2k_e$. At the onset of the
mixed phase $k_e$ and $N_{\boldsymbol{K}}$ are at a maximum, but as the
quark-hadron phase transition proceeds the negatively charged down and
strange quarks take over the process of charge neutralization.
Thus the electron number density and consequently $k_e$ continue to decrease
at about the same rate as before the start of the mixed phase.
This leads to the steep decline in $N_{\boldsymbol{K}}$ with
increasing $\chi$ for $\chi < 0.5$ shown in Figure
\ref{fig:lattice-vectors}. Further, the rod and slab dimensionality
drastically reduces the number of available scattering vectors which
contributes to the decrease of the MPB emissivity in those phases,
particularly in the slab phase. However, (\eqref{eq:j}) shows that the
MPB emissivity from rod and slab phases is dependent on $T^8$, rather
than $T^6$ for the blob phase, and this explains the dramatic decrease
in the MPB emissivity with decreasing temperature.

Direct Urca processes dominate the mixed phase neutrino emissivity at
all temperatures, with contributions from the
$\Lambda$ hyperon DU process ($\Lambda\rightarrow pe\bar\nu$)
operating beyond $\chi_{\mathrm{max}}$. Nucleonic DU processes do not
operate for any of the parametrizations considered in this work
\cite{Spinella2017:thesis}. The hyperon DU process
emissivities can be identified as any contribution with an emissivity
above that for the quark DU process in the $npe\mu$ composition, and
are shown to step down in the mixed phase, vanishing prior to the
onset of a pure quark phase. In the absence of the hyperonic DU
process the quark DU process would still dominate the Bremsstrahlung
and modified Urca processes unless curtailed by the presence of color
superconductivity. 

\section{Discussion and Summary}

In this work we determined that quark blob, rod, and slab structures
may exist in a crystalline quark-hadron mixed phase. The study is
based on relativistic mean-field equations of state which are used to
model hadronic matter and a nonlocal extension of the three-flavor
Nambu-Jona-Lasinio model for quark matter.  We determined the neutrino
emissivities that may result from the elastic scattering of electrons
off these quark structures (mixed phase Bremsstrahlung (MBP)), and
compared them to standard neutrino emissivity processes that may
operate in the mixed phase as well.

We found that the emissivity from the MPB process is comparable to
that of the modified Urca process at low volume fractions of quark
matter, $\chi$, and in the temperature range of $10^8 ~{\rm K} \lesssim T
\lesssim 10^9~\rm{K}$.  The MPB emissivity was found to increase with
the inclusion of the vector interaction among quarks and with
additional baryonic degrees of freedom in the form of hyperons and
$\Delta$ baryons \cite{Spinella2017:thesis}, both of which lead to an
increase in the quark-hadron phase transition density and a higher
density core.  Further, contributions to the MPB emissivity from
phonons were shown to be negligible compared to those from Bragg
scattering.  Finally, baryonic and quark DU processes were shown to
operate in the mixed phase and dominate all other neutrino emissivity
mechanisms.

Since it is believed that the hypothetical quark-hadron lattice
structures in the core regions of neutron stars are qualitatively
reminiscent to the hypothesized structures in the crustal regions of
neutron stars
\cite{Glendenning1992,Glendenning2001,Lamb1981,Ravenhall1983,Williams1985},
we have adopted the Bremsstrahlung formalism developed in the
literature for the crustal regions of neutron stars to assess the
neutrino emission rates resulting from electron-quark blob (rod, slab)
scattering in the cores of neutron stars with quark-hybrid
compositions. Because of the complexity of the problem, however, there
are several issues that need to be studied further in order to develop
refined estimates of the neutrino emission rates presented in this
paper. The remaining part of this section is devoted to this topic.

{\underline{Properties of the sub-nuclear crustal region:}} The
hypothetical structures in the crustal regions of neutron stars range
in shape from spheres to rods to slabs at mass densities $10^{14}
~{\rm g~cm}^{-3} \lesssim \rho \lesssim 1.5 \times 10^{14}~ {\rm
  g~cm}^{-3}$, which is just below the nuclear saturation density of
$2.5 \times 10^{14} ~{\rm g~cm}^{-3}$.  At densities where the nuclei
are still spherical in such matter, the chemical potential of the
electrons is $\mu_e \sim 80$ MeV and the atomic number of the nuclei
is $Z\sim 50$ \cite{Lorenz1993}. The corresponding Wigner-Seitz cell
has a radius of $R \sim 18$ fm, and the radius of the nucleus inside
the cell is $r \sim 9$ fm \cite{Lorenz1993}. The electrons moving in
the crystalline lattice formed by the ions are highly relativistic and
strongly degenerate. The ion coupling parameter, defined in
Eq. (\ref{eq:ion-coupling-parameter}), is $\Gamma \sim 2.3\times
10^{12} / T$, and the melting temperature $T_{\rm melt} \sim (Z e)^2 /
(R k_B \Gamma_{\rm melt})$ has a value of $T_{\rm melt} \sim 1.3
\times 10^{10}$~K.

{\underline{Properties of the quark-hadron lattice:}} The size of the
Wigner-Seitz cells associated with spherical quark blobs in the
crystalline quark-hadron phase is similar to the size of the
Wigner-Seitz cells in the crust. (Here, we do not consider the
crystalline phases made of quark rods and quark slabs since they
contribute much less to Bremsstrahlung because of the much smaller
number of electrons in those phases.)  For spherical quark blobs at
the onset of quark deconfinement, which occurs in our models at
densities of around three times nuclear saturation, $3 n_0$, the
electron chemical potential is $\mu_e = k_e \sim 140$ MeV. Hence, like
at sub-nuclear densities, the electrons are ultra-relativistic ($\hbar
k_e/m c^2 = 275$) and strongly degenerate. The electron degeneracy
temperature is around $T_F \sim 1.6 \times 10^{12}$ K, which is much
higher than the temperature range ($\lesssim 10^{10}$ K) considered in
this paper. From the results shown in Fig.\ \ref{fig:structure-radii},
one sees that the radii of the Wigner-Seitz cells containing spherical
quark-blobs are around $R\sim 12$ fm and that the quark blobs inside
the cells have radii of $r \sim 8$ fm.  The density of the
Wigner-Seitz cells is $(4 \pi R^3 /3)^{-1} \sim 1.4 \times 10^{-4}
~{\rm fm}^{-3}$ and the atomic number of the quark blobs inside the
Wigner-Seitz cell is around $Z \sim 200$.

{\underline{Plasma temperature and melting temperature:}} The ion
(quark blob) coupling parameter $\Gamma = (Z e)^2 / (R k_B T)$ is
given by $\Gamma = 6.7 \times 10^{13} /T$ and the melting temperature
of the ion crystal is $T_{\rm melt}= (Z e)^2 /(R k_B 172) \sim 4
\times 10^{11}$ K. Here we have used $\Gamma_{\mathrm{melt}} = 174$
for which a solid is expected to form \cite{Ogata1987,HaenselBook}.
Since the melting temperature of the quark crystal exceeds $10^{11}$ K
the quark blobs are expected to be in the crystalline phase at all
temperatures ($\lesssim 10^{10}$ K) considered in our study.  The
plasma temperature of the system follows from $T_P = 7.83 \times 10^9
\sqrt{Z Y_e \rho_{12} / A_i}$, where $Y_e = n_e/n_b$ is the number of
electrons per baryon, $n_e$ the number density of electrons, $n_b$ the
number density of baryons, and $\rho_{12}$ the mass density in units
of $10^{12}~{\rm g/cm}^3$.  For quark blobs with mass numbers of $A~
2000$, atomic number $Z \sim 200$, and $Y_e \sim 0.06$ one obtains a
plasma temperature of $T_P \sim 2 \times 10^{10}$ K.

{\underline{Electron-phonon scattering and Umklapp processes:}} In an
Umklapp process the electron momentum transfer in a scattering event,
$\hbar \vec q$, lies outside the first Brillouin zone, that is, $\hbar
q \gtrsim \hbar q_0$. This is in contrast to the normal processes
where $\hbar \vec q$ remains in the first Brillouin zone and $\hbar q
\lesssim \hbar q_0$, where $q_0 \approx (6 \pi^2 n_{\rm Blob}
)^{1/3}$. For the quark-blob phase we find $\hbar q_0 \sim 30$ MeV so
that $q_0/(2 k_e) \sim 0.13$ for the quark-blob lattice, which is of
the same order of magnitude as for the crust where $q_0/(2 k_e) = (4
Z)^{-1/3} \sim 0.01$ \cite{Kaminker1999}. The temperature below which
the Umklapp processes are frozen out is $T_{\rm Umklapp} \sim T_P
Z^{1/3} e^2 \sim 8 \times 10^8$ K, with the plasma temperature $T_P$
given just above. We find that the temperatures obtained for $T_{\rm
  Umklapp}$, $T_P$, and $T_{\rm melt}$ in the quark-blob phase are
rather similar to their counterparts in the nuclear lattice just below
nuclear saturation density, namely $T_{\rm Umklapp} \sim 10^8$ K, $T_P
\sim 10^9$ K, and $T_{\rm melt} \sim 10^{10}$ K .  In our study both
the Umklapp process and the normal process are taken into account
since temperatures in the range of $10^6 ~{\rm K}< T < 10^{10}~{\rm
  K}$ are considered.

{\underline{Debye-Waller factor:}} The effective interaction between
electrons and quark blobs depends on the thermal quark-blob lattice
vibrations which effectively smear-out the quark blob charges. This
feature is taken into account via the Debye-Waller factor given in
Eq.\ \eqref{eq:debye}. Since estimates for the Debye-Waller factor are
only known for spherical blob structures, the Debye-Waller may be the
largest source of uncertainly in our study.

{\underline{Role of electron band structure effects:}} It has been
shown in Ref.\ \cite{Pethick1994} that gaps in the electron dispersion
relation at the boundaries of Brillouin zones can noticeably reduce
the static lattice contribution. For point-like quark blobs with
atomic number $Z$ and for the smallest reciprocal lattice vector in a
bcc lattice, we estimate the electron band splitting from $0.018
(Z/60)^{2/3} k_e$ \cite{Pethick1994}. This leads to a splitting of
$\sim 6$ MeV for the quark-blob phase, which is around 1 MeV or more
for the nuclear lattice case \cite{Pethick1994}.

\acknowledgments{M.G.O.\ and G.A.C.\ thank CONICET and UNLP for financial support
  under Grants PIP 0714 and G 140. F.W.\ is supported by the National
  Science Foundation (USA) under Grants PHY-1411708 and PHY-1714068.}

\authorcontributions{The authors contributed equally to the
  theoretical and numerical aspects of the work presented in this
  paper.}

\conflictsofinterest{The authors declare no conflict of interest.}

\reftitle{References}

\end{document}